\documentclass[twocolumn,floatfix]{revtex4}

\usepackage{bm}
\usepackage{amsmath}
\usepackage{graphicx}
\usepackage{comment}

\begin{document}
\title{Metallic nanolayers -- a sub-visible  wonderland of optical properties}  
\author{A. E. Kaplan}
\email{alexander.kaplan@jhu.edu}
\affiliation{Dept. Chemical \& Biolog. Physics, Weizmann Inst. for Science, Rehovot 76100, Israel}
\affiliation{Dept. Electr. \& Comp. Eng., The Johns Hopkins University, Baltimore, MD 21218}

\date{\today}
\begin{abstract}
\noindent
It was predicted long ago that ultra-thin metallic films
must exhibit unusual optical properties
for radiation frequencies
from $rf$ to infrared domain.
A film would remain highly reflective
even when it is orders of magnitude thinner than
a skin depth at any frequency.
Only when it is a few nanometers thick
(depending on material but not on the frequency),
its reflectivity and transmittivity get equal, 
while its absorption peaks at 50\%.
It has been confirmed experimentally 
and new directions and applications were proposed.
We review the EM theory of the phenomenon and
recent developments in the field,
and present some new results.
$\copyright$ 2018 Optical Society of America
\vspace{.05in}

\end{abstract}

\maketitle
\vspace{-.1in}
\section*{Introduction}
\vspace{-.1in}
\noindent
The optics of metals is a prominent part of 
optical physics and related technologies,
in particular EM-devices from radio to microwave
to infrared wavelengths;
we will call it $sub-visible$ ($SV$) domain.
Of utmost importance to optics
is the capability of polished metal surface to 
serve as an almost ideal mirror, including  visible domain.
This quality was well known to
humankind from ladies mirrors in ancient Egypt,
to the legend of Archimedes' use of soldiers shields
to focus sunlight into the
enemy's ship sails, to radars, telescopes,
and other modern reflectors.

Due to very high conductivity of metals, $\sigma$,
the major properties of metallic mirrors 
in dielectric environment are:
(1) their reflectivity $R$
is very close to $1$ at any frequency $\omega$
within $SV$ domain and their absorption, $Q = 1 - R$, is 
respectively tremendously low,
(2) the electrical field at the reflecting
boundary nearly vanishes (i. e.
the reflected EM-wave is almost of the same
amplitude, but of the opposite phase,
forming thus a standing wave with a node at the surface);
(3) yet since the conductivity $\sigma$
is still finite, the exponentially 
decaying field penetrates into the metal
to a very shallow "skin depth", $\delta ( \sigma , \omega )$,
which is orders of magnitude lower than 
the wavelength of light in free space, 
$\lambda = 2 \pi c / \omega$, i. e. $\delta \ll \lambda$.
The phenomenon has been first experimentally 
explored by Hagen and Rubens [1], 
and electromagnetic theory for 
semi-infinite metallic layers was developed by Drude [2] 
more than a century ago.

Modern day applications and related physics 
call for the use of very 
thin metallic films, $d \ll \lambda $,
or even $d \ll \delta$ --
down to a few atomic layers -- and at the same time
offer capability of fabricating such thin films.
(The technique of making "gold leaves"
less than $0.1 \mu m$ thick was known to humans 
from the ancient times, and used in art and architecture [3].)
The issue arises then how thin must be such a layer to have
its reflectivity $R$ in $SV$
substantially reduced and its transmittivity, $P$, increased.
A characteristic thickness $d = d_{pk}$ would 
be say such that  $R = P$.
A common perception is that it happens when
the layer's thickness $d$ reaches
skin depth, $ d_{pk}  = O ( \delta )$,
and that the absorption gets even lower.

\vspace{-.01in}
A fact of the matter is that such 
a perception is wrong by orders of magnitude.
As was shown in [4], the reflectivity is 
drastically reduced only at amazingly small thickness, $d_{pk}$,
orders of magnitude lower than the skin depth,
down to a nanometer for metals,
corresponding actually  to
a small number of atomic layers.
Regardless of a specific metals, at such point
$R = P = 1/4$, and 
the absorption reaches its maximum,
$Q = 1 - R - P = 1/2$, very large as
compared to that of a semi-infinite layer.
(Furthermore, in counter-propagating waves,
a full absorption reaches $Q = 1$ [5]
at that point, and such a film becomes an 
ideal black-body, with $R = P = 0$.)
On the other hand,
easily fabricated films of the thickness $d < 0.01 \mu m$, 
i. e. greatly thinner than a skin depth, $d \ll \delta$,
may remain almost fully reflective,
$1 - R \ll 1$, so they can still
be used as good mirrors.

This thickness, $d_{pk}$,
is essentially a new and most
fundamental scale of metal optics in $SV$ domain,
as it is frequency-independent unlike $\delta$,
and relates only to $dc$ electronic properties of metal,
such as $dc$ conductivity (Sections 2),
or, under detailed consideration, - mostly
the density of free electrons (Section 6).
However, an amazingly simple nature of this effect
greatly overlooked in general literature,
is that at that thickness one attains         
impedance matching between the environment and
metallic layer resulting in maximum absorption.
In a free-space environment,
the $d_{pk}$-layer's impedance, $Z_{pk}$,
is exactly half of that of vacuum,
$Z_0 = 377 \ \Omega$, $Z_{pk} = Z_0 /2 $,
and it does not depend even on specific metal
(see details in Section 5 and 6).
It would be reasonable to call $d_{pk}$
an {\it{impedance-matching thickness}}.

The effect has by now been verified 
and explored both theoretically and experimentally.
That research included early [6 -- 9] $mw$ 
and recent [10 -- 13] millimeter wave experiments;
applications to the visualization of microwave modes 
using thermoluminescence sensors [14 –- 16], 
broadband millimeter wave spectroscopy  
in resonators at cryogenic temperatures [17 -- 19],
the theoretical and experimental study 
of  EM-properties of  periodic multilayers
of metallic films (photonic crystals) [20,21];
and a proposal to attain 100\% 
EM absorption in a standing wave [5].
However, it would be not an overstatement
to note that aside from those studies, 
this strongly pronounced 
and physically transparent effect remained 
little known to the research community in 
the optics of metals,
making it a blind spot in the field
(the choice of term "sub-visible" 
here is not accidental).
The theoretical and experimental
tools required for its exploration are not
overly sophisticated and were available 
for almost a century, yet even mentioning of it
is lacking not only from classical
texts on electrodynamics, 
but also from recent reviews on the subject.
The objective of this paper is to make a consistent
review of the major features of the optical properties 
of ultra-thin metallic films (including new results),
their underlying physics, and experimental results.

\vspace{-.01in}
A brief overview of optical properties
of semi-infinite metallic layers in $SV$ domain 
is found in Section 1.
Sections 2 and 3  are on the electrodynamics
and optical properties of the layers of finite thickness,
and Section 4 - on electrical currents.
Section 5 treats the problem in terms
of impedance theory, in particular
for arbitrary input/output environment.
Section 6 addresses the issue
of how the size dependent conductivity 
affects the optical properties of the layers,
and Section 7 - experimental results
and consideration for future experiments.
Section 8 is on the wave interference at metallic films
resulting in 100\%  absorption (blackbody effect);
and Section 9 briefly discusses potential applications
and outlook.
\vspace{-.05in}
\section*{1. Semi-infinite metallic layers}
\noindent
Major EM properties 
of semi-infinite (or sufficiently thick, $d \gg \delta$)
metallic layers, found 
in any "old goldies" texts, such as e. g.  [22-26],
can be represented by a succinct 
formula for the reflectivity, $R$,
and the absorption, $Q$,
of the layer for a normal EM-wave incidence [1]:
\begin{equation}
Q = 2 \delta k = \sqrt { 2 \omega / \pi \sigma} ; \ \ \
R  = 1 - Q ; \ \ \      P = 0
\tag{1.1}
\end{equation}
where $k = 2 \pi / \lambda = \omega / c$
is a wave number, $\lambda = 2 \pi c / \omega$
is a free-space wavelength for
an $\omega$-monochromatic wave,
$\sigma$ is a $dc$ conductivity of the layer
(we use here Gaussian units, 
see Appendix A, whereby  
$[ \sigma ] = s^{-1}$), and 
\begin{equation}
\delta = c / \sqrt {  2 \pi \sigma \omega } 
\tag{1.2}
\end{equation}
is a skin depth. 
For metals [and other highly conductive
materials, with $ \sigma \gg \omega / 2 \pi$ 
or $\lambda \gg c / \sigma$],
one has $\delta \ll \lambda$,
so that $Q = 1 - R \ll 1$;
this is due to large and almost purely
imaginary dielectric constant of metal,
$\epsilon_m \propto i \sigma / \omega$ (see Appendix A).
As an example, for a silver layer
at $\lambda = 1 m$, $\delta \approx 3.5 \mu m$
(respectively, for $\lambda = 1 cm$, $\delta \approx 0.35 \mu m$). 
Thus the EM-properties of the system, $R$, $Q$, and $\delta$
depend both on frequency of incident light $\omega$
and conductivity of the layer, $\sigma$, as expected.
Much less appreciated (yet known, see e. g. [24])
fact is that the total electrical current 
near metallic surface, $J$, induced by the incident
wave -- remains the same for any metal and wavelength,
and its amplitude, $J_{\infty}$, depends only 
on the incident amplitude $E_{in}$ as
\begin{equation}
J_{\infty} = c E_{in} / 2 \pi =  2 E_{in} / Z_0  
\tag{1.3}
\end{equation}
where $Z_0 = {4 \pi} / c$ 
is the free space wave impedance (see Appendix A).
A simple explanation of that is this.
Due to almost full reflection of light at the surface 
of a semi-infinite layer and  formation of standing wave 
in the free space with a node at the surface,
an $E$-field  there almost vanishes, while the magnetic
field peaks out, reaching amplitude $H ( x = 0 ) = 2 E_{in}$,
where $x$ is the distance from the surface.
Thus we have a rare situation
of an almost purely magnetic wave,
although it rapidly decays inside the layer,
$H ( \infty ) = 0$.
For a plane wave in a good metal (see details in Section 4 below), 
due to Amper's law, this $H$-field induces
local currents $j ( x )$,
$d H / dx = - j ( x ) Z_0$, see below, Eq. (2.2), so that
$ J_{\infty} = \int_0^{\infty} j ( x ) dx =
{H ( x=0 )  } / Z_0$, which results in (1.3).

It has to be noted though that
the ratio $E_{in} / J_{\infty} = Z_0 / 2$
is $not$ the impedance of the metal surface;
a respective impedance, $Z_{\infty}$, must relate the current $J_{\infty}$
to the $E$-field amplitude, 
\it at the surface%
\rm ,
$E_m (x=0) = E_{in} k \delta ( 1 - i ) $
(see Section (2) below). 
and $not$ to much larger amplitude of the incident wave, $E_{in}$,
so that
\begin{equation}
Z_{\infty} =  \frac{ E(x=0)}   {J_{\infty}} =  
\frac{Z_0}  2 k \delta ( 1 - i ) ;  \ \ \
Z_{\infty}  \ll  Z_0
\tag{1.4}
\end{equation}
Considering free space as a transmission line
for a plane wave, $ Z_{\infty}  \ll Z_0$
corresponds to its short-circuiting,
hence full reflection, as one would expect.
Rewriting (1.4) as
\begin{equation}
\frac 2 { 1 - i } \frac {Z_{\infty}} {Z_0 }= 
\sqrt { \frac {\omega}   {2 \pi \sigma }} =
\sqrt {k \Lambda } , \ \ \     with \ \ \
\tag{1.5}
\end{equation}
\vspace{-.2in}
\begin{equation}
\Lambda = \frac c {2 \pi \sigma } = 
k \delta^2 = \frac 2   { Z_0 \sigma } = O ( 1 ) {\AA} 
\tag{1.6}
\end{equation}
we introduce a new, frequency independent
characteristic length scale of a layer, $\Lambda$,
which will become one of the major "characters"
of the story for very thin layers.
It is a characteristic scale at which,
{\it is one presumes that the layer conductively,
$\sigma$ remains the same as the bulk
conductivity $\sigma_0$,}
the reflection would gets significantly
reduced, and the transition respectively increased.
For good metals, this new scale has atomic size
(and even less than that)
and is not only many orders of magnitude
smaller than the  wavelength of incident light,
but also of the skin depth.
In real layers, however, the conductivity,
$\sigma$ depends of the layer thickness, $d$, 
and gets greatly reduced as $d \to 0$
(due to the fact the mean free path of electrons, $l$, 
with $\sigma \propto l$, gets "clipped" by the walls).
see Section 6  below.
This "clipping" results in
the formation of another scale, $\lambda_N$,
which will finally determine the 
depth, $d_{pk}$, at which 
the layer will universally 
to the point where the reflectivity, $R$,
will be exactly equal to the
transmittivity, $P$, with $R = P = 0.25$,
and absorption will peak at $Q = 0.5$,
see Fig. 1 for the case of silver layer
and normal accidence.
\begin{figure} [h]
\begin{center}
\includegraphics[width=5cm, height=8cm, angle=270]{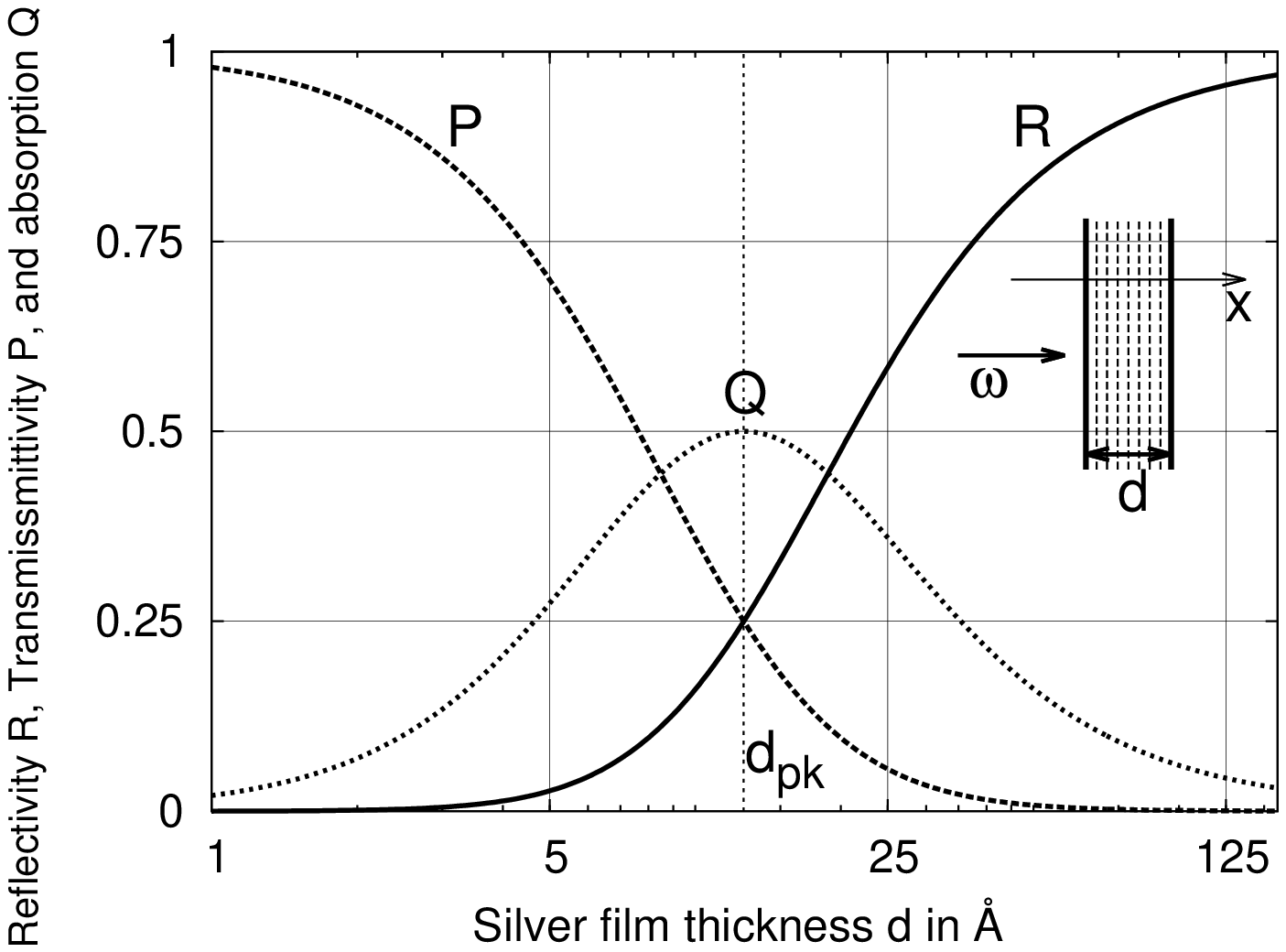}
\caption{Theoretically calculated refractivity, $R$,
transmittivity, $P$, and absorption, $Q$,
of a silver layer $vs$ its thickness $d$;  
$d_{pk} \approx 12.6$ {\AA} [4]. 
Insert: a normal plane wave incidence.}
\label{fig1}
\end{center}
\vspace{-0.3in}
\end{figure}
This new scale is of the order of $1 - 2 \  nm$,
and defined as $\lambda_N =$ $ \sqrt 
{\Lambda_0 l_0} \approx 8.2 \times N_e^{-1/3}$,
where $\Lambda_0 = \Lambda ( d \to \infty )$,
$l_0 = l ( d \to \infty )$ 
is the bulk mean free path of electrons,
and $N_e$ is the number 
density of free electrons
(for details, see Section 6).

The conductivity $\sigma$ in $SV$ domain
remains essentially the same as for a $dc$  current [1,2],
i. e. the entire phenomenon is of a (quasi)static nature.
This greatly simplifies the theory
of the optical properties of metal 
in particular thin metal films, in that domain,
and makes the entire $SV$ domain so special.
However, at the upper-energy end of this domain,
the interaction of radiation with quasi-free electrons, 
at least in semi-infinite layers,
reaches the point where
the skin depth gets smaller than 
the mean free path of electrons, $l_0$,
which results in the so called
anomalous skin effect [27].
The interaction then becomes non-local:
the electrons driven by the field near
the surface, run away into a no-field area.
For example, in the case of silver, 
the wavelength at which this happen
is in sub-mm domain, $\lambda  \sim 0.23  mm$.
For shorter wavelength radiation,
an absorption is increasing;
however the metals mirrors 
still remain well-reflecting even 
in the visible domain.
However, for thin layers (see Section 6), the onset of
non-locality shifts to shorter wavelengths, since
$l_0$ gets closer to the layer thickness $d$;
quasi-static model of optical constants of nano-layers
holds true to the mid-infrared domain.
A review of the electronic properties
of various metals and their related optical properties
from classical Drude-Lorentz model to
the quantum theory for various metals
and frequencies can be found in [28 -- 31].

For the higher photon energies, as e. g.
in $UV$ spectral domain,
the quasi-free electrons can be regarded
as an over-dense plasma (see also Appendix A),
having now almost real 
yet negative dielectric constant 
$\epsilon \approx 1 - \lambda^2 / \lambda_{pl}^2 < 0$.
with the plasma wavelength $\lambda_{pl}$
found in $200 - 300  / n m$ range
determined by the density of electrons.
Beyond that threshold, metals can be viewed
as a plain under-dense plasma,
with its dielectric constant 
approaching that of a free space, $0 < \epsilon < 1$.
Further into X-ray domain,
the number of free electrons undergoes large 
jump-like increases as photon energy increases
near so called $K$, $L$, $M$, and $N$ absorption edges, 
which are due to resonant photo-ionization of bound
electrons from respective shells into conduction state [32--34].
These jumps would affect the optical constants of metal surface,
and may be used for various applications,
in particular for narrow-line transition radiation
by electron beams traversing a multi-layer,
super-lattice structure of
metallic layers generating almost coherent
radiation in soft [35] and hard X-ray [36] domains.
In this paper we limit our consideration
only to the quasi-static, i. e. $SV$ spectral domain.
\vspace{-0.3in}
\section*{ 2. Fields in a layer of finite thickness }
\vspace{-0.1in}
\noindent
The optical properties of thin metal layers
(both in visible and far-infrared domain) 
with a lot of experimental  data
has been reviewed in [37--40]
(a relevant research has been done on absorption
by small metal particles in infra-red, [41]).
Some of them clearly pointed to 
layers ability to sustain high reflectivity
at surprisingly small thickness 
(see e. g. [42,43]); yet a general
picture of existence of universal
maximum absorption and a corresponding
spatial scale related only to the number
density of free electrons (Section 6 here)
did not seem to transpire yet.
It is also interesting to note that 
a tremendous amount of work
on the physics of superconductivity,
especially on high-TC superconductivity,
has been done by studying infrared and
far-infrared optical properties
of thin films of those materials
(see e. g. review [44]),
with some of them emphasizing 
that the spatial scale of
strongly absorbing films
are below the skin depth (see e. g. [45]).

This and the following Sections 
are to provide a basic
understanding of how the ultra-thin
metallic films reaches the state
of maximum absorption, $Q = 0.5$,
and to show that this behavior is really universal.
While looking for the fields and current 
in a (non-magnetic) layer of finite thickness,
for our purposes, we consider only normal incidence, 
i. e. the waves propagating in the $x$-axis
normal to the layer, Fig. 1.
In this section, we consider only a free space
as an environment, where the layer 
is bounded between $x = 0$, and $x = d$.
(The results are extended to arbitrary
environment in Section 5.)
All the calculations in this paper are based on most
simplifying assumptions,
sufficient to bring up and quantitatively describe major 
features of the phenomenon discussed:
the layer has flat surfaces, and
the metal in it is homogeneous (i. e. not granulated),
so that there is no scattering of light at the layer;
the conduction electrons are regarded
as a gas of non-interacting
particles following Drude model [46],
that may get scattered by ions and 
the surfaces of the layer.
An $\omega$-monochromatic
wave is incident normally at the layer
at the point $x = 0$.
(For the incidence close to the normal
one, the results are expected to be
not much different, since the 
refractive index of metals is large.)
The wave is linearly polarized in the $y$-axis,
both incident fields $(1/2) E (x) e^{{-} i \omega t} + c. c.$ and 
$(1/2) H (x) e^{{-} i \omega t} + c.c.$
have the same amplitude $E_{in}$, and
$E_z = H_y = E_x = H_x = 0$ everywhere.
We designate $E \equiv E_y$, and $H \equiv H_z$.
In this case, the Eqs. (A.1) and respective wave equation
e. g. for $E$ are written as
\begin{equation}
\frac {dE }  {dx} = i k H  ; \ \ \ \
\frac {dH } {dx} = - i \epsilon k {E } ;  \ \ \ 
\frac {d^2 E}  {d x^2 } + \epsilon k^2 E = 0 .
\tag{2.1}
\end{equation}
where $k = \omega / c = 2 \pi / \lambda$
and $\lambda = 2  \pi c / \omega $
are free-space wavenumber 
and wavelength of the wave respectively.
The free-space 
dielectric constant 
is $\epsilon = 1$, whereas inside the 
layer, $0 \le x \le d$, 
under a "good metals" condition,
$ \epsilon_m  \gg 1$, or $ \sigma \gg \omega$,
$\epsilon_m$ can be well 
approximated as purely imaginary quantity
(see also Appendix A):
\begin{equation} 
\epsilon_m \approx \frac {4 i \pi \sigma}  {\omega }= 
\frac {2i} {k \Lambda } = \frac {2i} {k^2 \delta^2 } ,
\ \ \ \
\Lambda  = \frac c {2 \pi \sigma }
\tag{2.2}
\end{equation}
where a scale $\Lambda$ [4] 
was introduced in (1.6),
and $\delta$ is as defined in (1.2).
The incident $E$-field of amplitude $E_{in}$, 
$H$-field of amplitude $H_{in}$,
and reflected  fields
of amplitude $E_{rf}$ and $H_{rf}$
at $x < 0$ are respectively as
\begin{equation}
E (x) / E_{in} = H (x)  / E_{in} = e^{ i x k } 
\notag
\end{equation}
\vspace{-.3in}
\begin{equation}
E_{rf} (x) / E_{in} = 
- H_{rf} (x) / E_{in}  =
r e^{ - i x k }
\tag{2.3}
\end{equation}
The transmitted fields behind the layer, 
i. e.  at $x > d$ will be sought for as
\begin{equation}
E_{tr} / E_{in}  =  H_{tr} / E_{in}  =
p e^{ i ( x - d ) k }
\tag{2.4}
\end{equation}
where $r$ and $p$ are the coefficients or reflection
and transmission respectively
to  be found from boundary conditions
at the surfaces of the layer.
(The solution (2.4) for the transmitted field 
takes into account the so called Zommerfield's
condition in the infinity ($x \to  + \infty$),
by ruling out a back-propagating wave 
$\propto  e^{ - i x k }$ at $x > d$.)
Inside the layer the fields
$E_m$ and $H_m$ are superpositions
of forward ("+") and backward ("-")
propagating waves with
normalized amplitudes $a^{\pm}$ as
\begin{equation}
{E_m (x)} /  {E_{in} }= a^+  e^{{i} k_m x } +
a^- e^{{-} i k_m x } ;      
\notag
\end{equation}
\vspace{-.3in}
\begin{equation}
{H_m (x)  /  E_{in} }= n_m  \left( a^+ e^{{i} k_m x } -
a^- e^{{-} i k_m x } \right) 
\tag{2.5}
\end{equation}
with
\begin{equation}
k_m = k \sqrt \epsilon_m = 
\frac {1 + i} {\delta } = k n_m ; \ \ \
n_m = \sqrt \epsilon_m = \frac {1 + i} { k \delta}
\tag{2.6}
\end{equation}
where $n_m$ is a (complex)
refractive index of metal.
The constants $a^{\pm}$ are found 
from boundary conditions at $x=0$ and $x=d$
for $E$ and $H$ (to be continuous at a boundary).
Using (2.4), (2.5), we have at $x=0$:
\begin{equation}
a^+ + a^- = 1 + r ;     \ \ \ \ 
n_m ( a^+ - a^- ) = 1 - r
\tag{2.7}
\end{equation}
and at $x = d$:
\begin{equation}
a^+ e^{{i} k_m d } +
a^- e^{{-} i k_m d } = 
\notag
n_m \left( a^+ e^{{i} k_m d } -
a^- e^{{-} i k_m d } \right) = p
\end{equation}
From these equation,
the solution for $a^{\pm}$ is then
\begin{equation}
a^{\pm} =
- \frac {2 (  n_m  \pm 1 ) e^{\mp i k_m d }} 
{( n_m  -  1  )^2  e^{i k_m d } - ( n_m  +1 )^2  e^{-  i k_m d } }
\tag{2.8}
\end{equation}
Using the fact that $ \epsilon_m  \gg 1$,
hence $n_m \pm 1
\approx n_m e^{{\pm} 1 / n_m}$,
we simplify (2.8) as
\begin{equation}
a^{\pm} \approx  \frac 1 {n_m }
\frac  {e^{\pm ( 1 / n_m - i k_m d ) }}  
{\sinh (  2 / n_m - i k_m d  ) }
\tag{2.9}
\end{equation}
For very thin layers, $d \ll \delta$, we have 
\begin{equation}
a^+ \approx   a^- \approx 
( {\Lambda / 2 ) / ( \Lambda + d } ) = const ;
\tag{2.10}
\end{equation}
(for $E(x)$ see below, (2.14)),
i. e. the both counter-propagating waves,
are of almost the same amplitude.
Eq. (2.5) yields now for the
fields inside the layer:
\begin{equation}
\frac {E_m (x)}  {E_{in} } \approx
\frac 2 {n_m } 
\frac { \cosh [ i k_m ( x - d ) + 
1 / n_m ]}   {\sinh ( 2 / n_m - i k_m d ) } ;       
\notag
\end{equation}
\begin{equation}
\frac {H_m (x)}  {E_{in} } \approx
2 \frac { \sinh [ i k_m ( x - d ) + 1 / n_m  ]  } 
{\sinh ( 2 / n_m - i k_m d ) }
\tag{2.11}
\end{equation}
hence
\begin{equation}
{H_m (x)  / E_m (x)}  \approx
n_m \tanh  \left[ i k_m ( x - d ) 
+ {1 / n_m }\right] 
\tag{2.12}
\end{equation}
in particular, for a semi-infinite layer, $d \to \infty$, 
\begin{equation}
{H_m (x)} /  {E_m (x)}  = const = n_m 
\tag{2.13}
\end{equation}
In most interesting case of 
a thin layer, $d \ll \delta $, we have
\begin{equation}
\frac {E_m (x)}  {E_{in}} \approx 
\frac {\Lambda} {\Lambda + d } ;
\ \ \ \
\frac {H_m (x)}  { E_{in} } \approx
\frac { \Lambda  + 2 ( d -  x )}   {\Lambda  + d }
\  \   \
\tag{2.14}
\end{equation}
i. e.  $E_m (x) / E_{in} = const$,
so the electrical field $E_m$ is homogeneous
inside the layer.
\vspace{-.1in}
\section*{3. Reflectivity, transmittivity, and absorption}
\noindent
From (2.7) $r = ( a^+ + a^- ) - 1$, hence, using (2.8) 
\begin{equation}
r = - \frac { (  \epsilon_m  - 1 ) ( e^{{i} k_m d } -
e^{{-} i k_m d } )}   
{( n_m  -  1  )^2   e^{{i} k_m d } - ( n_m  +  1  )^2 e^{{-}  i k_m d } } 
\tag{3.1}
\end{equation}
If $d = 0$, we have $r =  0$, as expected.
If $d \to \infty$, the terms
$e^{{i} k_m d }$ tend to zero, so that
\begin{equation}
r_{d \to \infty} =  - 
( n_m  - 1 ) /   ( n_m  + 1 )
\approx
-  1 + { k \delta } ( 1 - i )       
\tag{3.2}
\end{equation}
and consistently with (1.1) we have
\begin{equation}
R_{\infty} =  |r|^2 = 1 - 2 k \delta
\tag{3.3}
\end{equation}
For $ \epsilon_m  \gg 1$,
similarly to (2.9), (3.1) is simplified as
\begin{equation}
r \approx -  
\{ \sinh [ ( 1 - i ) d / \delta ] \} /
\{ {\sinh [  ( 1 - i ) ( d / \delta + k \delta ) ] } \}
\tag{3.4}
\end{equation}
In the same way, we have
the transmission coefficient, $p$:
\begin{equation}
p = [ (1 - i ) k \delta ] /
\{ sinh [ (1 - i ) ( k \delta + d / \delta ) ]   \}
\tag{3.5}
\end{equation}
Both of them can be further simplified
for the case of very thin layer, $d \ll \delta$ [4]:
\begin{equation}
r \approx  - d /  ( d + \Lambda ) , \ \ \ \
p \approx  \Lambda /   (  d  +   \Lambda )
\tag{3.6}
\end{equation}
Translating Eqs. (3.6) 
for thin layers into the formulas for
reflectivity, $R =  |r|^2 $,
transmittivity, $P =  |p|^2$,
and energy losses, $Q = 1 - ( R + P )$, 
we get:
\begin{equation}
R = \left( 1 + {\Lambda } / d \right)^{-2} ;    
\ \ \ \
P = \left( 1 + {d / \Lambda }\right)^{-2} ;    
\notag
\end{equation}
\vspace{-.3in}
\begin{equation}
Q = 2  \left( \sqrt { {\Lambda / d }} + 
\sqrt { {d / \Lambda }} \right)^{-2}
\tag{3.7}
\end{equation}
Interestingly enough, (3.7) provides us with
a relationship that doesn't 
explicitly include any parameters
of the incident field or the metal:
\begin{equation}
\sqrt R + \sqrt P = 1 ;      
\ \ \ \ \
Q = 2 \sqrt { R P} 
\tag{3.8}
\end{equation}
Notice that the first of these equations,
written as $p = 1 + r $ 
is not related to the conservation of full
momentum in the system;
indeed that conservation should
include the momentum, $p_m$,
transferred to the layer:
\begin{equation}
p_m = 1 - p - r = 2 d / ( d + \Lambda )
\tag{3.9}
\end{equation}
which originates a radiation pressure on the layer.
In the semi-infinite layer case, $p_m \approx 2$,
i. e. is maximal, as expected, 
whereas $p_m = 0$ at $d = 0$, and finally
$p_m = 1$ at $d = \Lambda $.
However, (3.7) and (3.8) uphold
the conservation of radiation energy, $R + P + Q =1$.
\vspace{-.1in}
\section*{ 4. Electrical current in the layer}
\noindent
As long as the solution for electrical and magnetic 
fields (2.11), (2.14) are known,
the electrical current $j (x)$ in the layer is found as 
\begin{equation}
j (x) =  - Z_0^{-1} {\partial H } / {\partial x} 
= \sigma E ;
\tag{4.1}
\end{equation}
For a thin layer, $d \gg \delta$,
the field $E$ is almost constant (2.14),
so the current $j (x)$ is also
evenly distributed along  the depth.
The $full$ current, $J (d) $, in the layer is 
\begin{equation}
J (d) = \int_{x=0}^{x=d} j (x) dx = 
\frac { H_m (x=0) - H_m (x=d)}   {Z_0 }
\tag{4.2}
\end{equation}
or by using the second equation in  (2.11), we have
\begin{equation}
\frac {J (d)}  {E_{in} }\approx
\frac 2 {Z_0 } \ \ \frac  {  \sinh ( i k_m d / 2 ) } 
{\sinh ( i k_m d / 2 - 1 / n_m ) } 
\tag{4.3}
\end{equation}
For a very thin layer, $d \ll \delta$, we have
\begin{equation}
\frac {J_d} { E_{in} }= 
\frac 1 {Z_0} \frac d {d + \Lambda } = \frac {|r|}  { Z_0 } ;
\tag{4.4}
\end{equation}
In a semi-infinite layer, $d \gg \delta$, or $k_m d \gg 1$,
we have from (4.3) (see also (1.3)):
\begin{equation}
{J ( d \to \infty )   / E_{in} }= 2 / Z_0 ;    
\tag{4.5}
\end{equation}
As long as $d > \Lambda$, the total
current is almost the same!
Furthermore, the efficient layer resistance per square cm,
coincides exactly with the half-impedance of vacuum.
Thus the layer almost always makes the
same radiating antenna, from $d \gg \delta$ down to $d \sim \Lambda$.
\vspace{-.1in}
\section*{5. Transmission line analogy; arbitrary environment}
\vspace{-.1in}
\noindent

It is instructive and revealing to describe 
EM wave propagation through a thin metallic layer
as a transmission line problem, 
by using wave impedances of the line and its components.
A simple impedance algebra 
allows for an easy generalization of our results
to a system with arbitrary (i. e. not just free space)
input/output environment,
which may include e. g. dielectric or 
semiconductor substrate.
At the same time if offers
a plain "electrical engineering" view
of the phenomenon.

Let us start with a free space environment.
Both of the impedances of the "incident"
and "output" arms of the line is then $Z_0$, 
and in view of quasi-static nature of the problem,
a layer can be regarded as a lump circuit
(even for a semi-infinite layer),
which in the case of $d \ll \delta$ is simply a resistor.
Thus, for all the purposes,
using (2.11) with $x=0$ and (4.3),
the impedance $Z_L$
of a layer of a thickness $d$ is found as
$Z_L  (d) =  {E_m (x=0)} /  {J (d)}$, or
\begin{equation}
Z_L  (d) \approx  - \frac {( Z_0 / 2n_m ) \cosh ( i k_m d - 1 / n_m  )}
{\cosh ( i k_m d /2 - 1 / n_m ) \sinh ( i k_m d / 2 ) } 
\tag{5.1}
\end{equation}
In the limit $d \to \infty$ we have
\begin{equation}
Z_{\infty} / Z_0  =
n_m^{-1} =
k \delta ( 1 - i ) / 2 
\tag{5.2}
\end{equation}
We can also find $Z_{\infty}$ directly from the wave 
solution (2.13).
Indeed, for a plain $travelling$ wave, a wave impedance $Z$
in electrodynamics is usually defined as
\begin{equation}
Z = \left( E / H \right)_{SI} = Z_0 \left( E / H \right)_{Gauss} ,
\tag{5.3}
\end{equation}
where subscripts refer to a respective unit system.
This is still true even if 
the wave goes through an absorbing material,
if there is no retroreflection inside it.
In a semi-infinite metallic layer,
the ratio $E / H$ remains constant, (2.13), 
since the wave propagate only away from
the interface (see e. g. (2.5) with $a^- = 0$ as
follow from (2.9) with $d \to \infty$), and thus we have 
$Z_{\infty} / Z_0 =$ $ E_m (x) / H_m (x) = 1 / n_m $
which coincides with (5.2).
In the limit  $d \ll \delta$ we have
for the layer impedance $Z_L$:
\begin{equation}
Z_L  / Z_0 \approx 
- 1 /  i \epsilon_m k d  =
\Lambda / 2d 
\tag{5.4}
\end{equation}
The coefficient of reflection, $r$, of the layer, if
the wave is incident from $x \to  - \infty$,
can be evaluated by assuming that the incidence
line with $Z_{in} = Z_0$ is 
loaded by the impedance $( Z_{\Sigma} )_0$ formed by
two elements connected {\it in parallel}:
the layer, with its impedance $Z_L $,
and the output line with its impedance $Z_0$, i. e.
\begin{equation}
( Z_{\Sigma} )_0^{-1} = Z_L^{-1} + Z_0^{-1}
\tag{5.5}
\end{equation}
so that a transmission line theory yields
\begin{equation}
r = \frac  {( Z_{\Sigma} )_0 - Z_0}
{( Z_{\Sigma} )_0 + Z_0} =
-  \frac {Z_0} {Z_0 + 2 Z_L } ; \ \ \ \    
R = | r |^2
\tag{5.6}
\end{equation}
Similarly, the coefficient of transmission $p$ is evaluated as
\begin{equation}
p =  \frac { 2 ( Z_{\Sigma} )_0}   
{( Z_{\Sigma} )_0 + Z_0 } =    
\frac {  2 Z_L}   {Z_0 + 2 Z_L } ;  \ \ \ \      
P =  | p |^2 
\tag{5.7}
\end{equation}
so that for $d \ll \delta$ they coincide with the 
respective result (3.6).
Finally, the energy losses are as
\begin{equation}
Q = 1 - ( P + R ) = 
{ 4 Z_0 Z_L} /   {( Z_0 + 2 Z_L )^2 } 
\tag{5.8}
\end{equation}
which peaks ($Q_{pk} = 0.5$) at
$Z_L = Z_0 / 2$, as expected.

The transmission line results
can be readily generalized to the case whereby
semi-infinite dielectric materials sandwiching
the metallic layer are {\it different}.
Assuming that the material of the wave incidence
has a refractive index $n_1$ and the output one -- 
the index $n_2$, so that their respective
wave impedances are $Z_i = Z_0 / n_i$ with $i = 1,2$,
we define "input/out load impedance" $Z_{\Sigma} $ as
\begin{equation}
{ Z_{\Sigma}^{-1} } =  Z_L^{-1} + Z_2^{-1} ,
\tag{5.9}
\end{equation}
and use it to generalize (5.6) and (5.7) as:
\begin{equation}
r =  \frac {Z_{\Sigma}  - Z_1}  {Z_{\Sigma} + Z_1} = -  
\frac {  Z_1 Z_2  +  Z_1 Z_L - Z_2 Z_L } 
{Z_1 Z_2 + Z_1 Z_L + Z_2 Z_L  } = 
\notag
\end{equation}
\vspace{-.2in}
\begin{equation}
- \frac  {  Z_0 / Z_L  +  n_2 - n_1  }
{Z_0 / Z_L  +  n_2 + n_1 } ;  \ \ \ \  
R = | r |^2 ;
\tag{5.10}
\end{equation}
\vspace{-0.2in}
\begin{equation}
p =  \frac { 2 Z_{\Sigma}}   {Z_{\Sigma} + Z_1 } =
\frac {  2 Z_2 Z_L }  {Z_1 Z_2 + Z_1 Z_L + Z_2 Z_L  } =
\notag
\end{equation}
\vspace{-.2in}
\begin{equation}
2 n_1 /   ( Z_0 / Z_L  +  n_2 + n_1 ) ;  \ \ \ \ 
P = ( n_2 / n_1 ) | p |^2
\tag{5.11}
\end{equation}
We obtain then the energy losses as
\begin{equation}
Q = 1 - ( R + P ) =
\frac { 4 Z_L Z_2^2  Z_1 }   
{ (  Z_1 Z_2 + Z_1 Z_L + Z_2 Z_L )^2 } =
\notag
\end{equation}
\vspace{-.2in}
\begin{equation}
4 n_1 ( Z_0 / Z_L )   /   ( Z_0 / Z_L  +  n_2 + n_1 )^2 ;
\tag{5.12}
\end{equation}
If $n_1 = n_2 = 1$, the results 
(5.9)-(5.11) coincide with (5.7) - (5.8),
including the case $d \ll \delta$, 
when they coincide with (3.7).
The losses (5.12) then reach their maximum 
\begin{equation}
Q_{pk} = \frac {n_1} {n_1 + n_2 } \ \ when \ \
\frac {Z_0} {Z_L } \left( =  \frac {2d_{pk}}  {\Lambda }\right)  =
n_1 + n_2   
\tag{5.13}
\end{equation}
while the reflectivity $R$ and transmittivity $P$ are as
\begin{equation}
R = \frac {n_2^2}  {( n_1 + n_2 )^2 },    \ \ \ 
P = \frac {n_1 n_2}   {( n_1 + n_2 )^2} ,  \ \ \
\frac R P = \frac {n_2} {n_1 }
\tag{5.14}
\end{equation}
Note that if in (5.13) $n_1 \gg n_2$,
the losses $Q$ greatly increase;
this may happen if dielectric
at the entrance is highly optically dense,
as in some semiconductors,
or if the output medium consists of plasma
near critical frequency, when $0 < n_2 \ll 1$.
For example, if
one uses silicon ($Si$), or gallium arsenide ($GaAs$),
both of which have refractive index $n \sim 4$,
as an input medium ($n_1 $),
and an air as an output one ($n_2 = 1$),
one  would have highly absorbing and low reflecting layer 
$Q_{pk} = 0.8$, 
$R = 0.04$, and $P = 0.16$ at 
$d_{pk} / \Lambda = 2.5$.

Having in mind that e. g. in a free
space the maximum absorption happens
when the layer impedance matches
exactly half of vacuum impedance,
regardless of the specific material,
it might be perhaps appropriate to call
the entire phenomenon an {\it impedance-match absorption}.
\vspace{-.2in}
\section*{6. Size-affected conductivity,
and new fundamental thickness scale}
\noindent
For ultra-thin films, eq. (3.7) suggests
a very simple and transparent
dependence of optical properties $R$, $Q$, and $P$ $vs$
film thickness $d$ -- {\it assuming that
the conductivity $\sigma$ and the
scale $\Lambda =  c / 2 \pi \sigma $ (2.2)
are constant that don't depend on $d$}.
(The resulting scale $\Lambda$ is around or even 
less than one angstrom.)
But at very small $d$ this is not true anymore,
so that $\sigma$ $vs$ $d$ dependence 
has to be taken into consideration.
The major parameter through which
the specific conductivity of good metal
is affected by the film size, $d$,
is the mean free path of electrons, $l (d)$,
with the conductivity $\sigma$
being proportional to $l (d)$ [47,48]:
\begin{equation}
\frac {\sigma ( d )}  {\sigma_0 } = 
\frac {l (d)}  {l_0 };    \ \ \ with \ \ \
\sigma_0 =
\frac {N_e e^2 l_0}   {\sqrt { 2 m W_F } } ,
\tag{6.1}
\end{equation}
where $\sigma_0$ and $l_0$
are respectively specific bulk
conductivity and bulk mean free path,
$N_e$ is number density of conduction electrons,
$m$ is the electron mass,
$W_F$ is the Fermi energy of an electron gas 
at a given temperature;
for the most applications
the temperature are less then $10^3 - 10^4 K$,
so $W_F$ is the same as for absolute zero [49],
$W_F \approx W_0 = ( \hbar k_F )^2 / 2 m$,
where $k_F = ( 3 \pi^2 N_e )^{1/3}$ 
is the Fermi wave vector,
which reduces $\sigma_0$ in (6.1) to
\begin{equation}
\sigma_0 = {\alpha k_F^2 c l_0} /   {3 \pi^2 } ;  \ \ \
with \ \ \
\alpha = {e^2 / \hbar c} = {1 / 137 }
\tag{6.2}
\end{equation}
where $\alpha $ is the fine structure constant.
The experimental and theoretical data for $l_0$
can be found in many publications;
the latest extensive study for 
20 metals using numerical calculations
over the Fermi surface found in [50].
The size effect, i. e. how $\sigma $
and $l$ depend on $d$
represents fundamental interest as well as
application challenges for nano-electronics.
Sufficient for our purposes here is 
a classical Fusch-Sondheimer model [47,48],
which assumes a so called a spheric Fermi surface,
whereby  the major factor affecting
$\sigma$ and $l $ is electron scattering
at the layer surfaces.
Within that model, under the most realistic 
assumption that electrons scatter
at the surface in a purely diffuse way, 
the dimensionless mean-free path of electrons
and the conductivity,
$\theta =l ( d ) / l_0 = {\sigma ( d )} /  {\sigma_0 }$ 
$vs$ the layer width, $\xi = d / l_0$ is as [48]:
\begin{equation}
\theta =
1 - \frac { 3 }   {2 \xi}  \int_1^{\infty} 
\left( t^{-3} - t^{-5} \right) 
\left( 1 - e^{{-}  t \xi } \right)  d t
\tag{6.3}
\end{equation}
%
or in terms of  exponential integral 
$E_1 ( \xi ) = \int_{\xi}^\infty 
\frac {e^{-  t}} t dt$ [51]
\begin{equation}
\theta =
1 - 3 ( 1 - e^{- \xi} ) /  {8 \xi}  + 
\notag
\end{equation}
\vspace{-.3in}
\begin{equation}
[ ( - 10 - \xi + \xi^2 ) e^{- \xi} +
\xi ( 12 - \xi^2 ) E_1 ( \xi ) ] / 16
\tag{6.4}
\end{equation}
In the limit of very thick 
layer, $\xi \gg 1$, the solution (6.4)
is $\theta \approx 1 - 3 / {8 \xi } $,
whereas in the limit of
very thin layer, $\xi \ll 1 $,
which is of  most physical interest,
the solution is
$\theta \approx \left( {3 \xi} /  4 \right) ln 
\left(  1 / {\xi } \right)$.
\begin{figure} [ht]
\begin{center}
\includegraphics[width=4.5cm, height=7cm, angle=270]{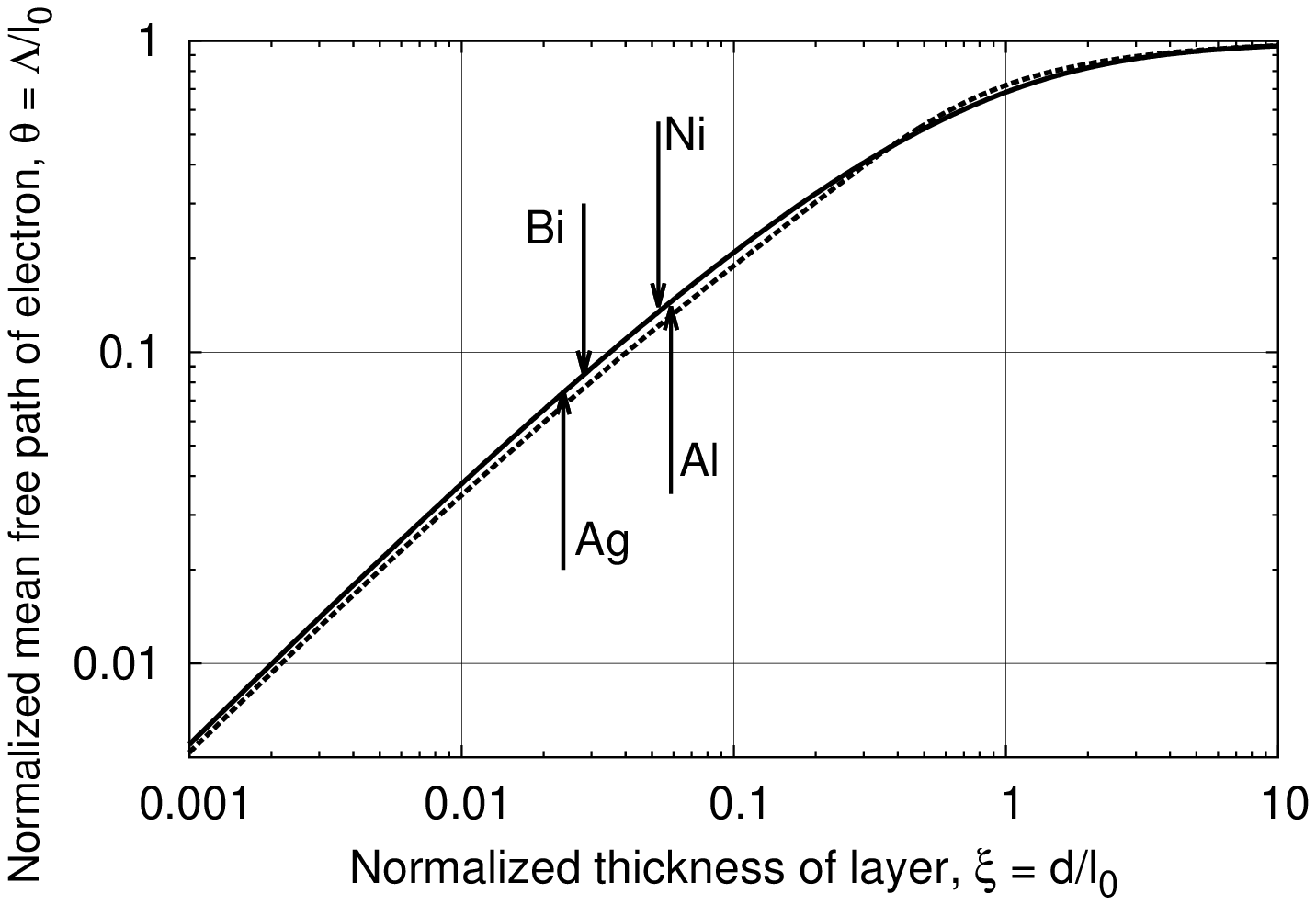}
\caption{Normalized mean free path of electrons,
$\theta = \Lambda ( d ) / l_0$, $vs$
normalized thickness of layer,
$\xi = d / l_0$.
The upper curve is due to exact solution,
Eq. (6.4), the lower one - due to interpolation (6.5).
Arrows indicate values $\xi_{pk}$ and $\theta_{pk}$
of peak absorption, $Q = 0.5$
based on Table 1, for selected metals.}
\label{fig2}
\end{center}
\end{figure}
To simplify the analysis of the result (6.4),
we found that for all the practical purposes,
in particular in the above limits, it is 
interpolated with great precision
within an entire range $0 < d < \infty$ by 
\begin{equation}
\theta = 
( 4 + \xi^{-2})^{-1} +
( {3 \xi} /  {4}) 
ln \left( 1 + \xi^{-1} \right)      
\tag{6.5}
\end{equation}
see the comparison of (6.4) and (6.5) at Fig. 2.
A layer thickness, $d_{pk}$, where
the absorptions peaks out, $Q_{pk} = 0.5$,
according to (3.7), satisfy the condition
$\Lambda (d_{pk}) = d_{pk}$.
Thus to solve (6.5), we 
recall that ${l (d)} / {l_0} =$
$ {\Lambda_0 } /   {\Lambda (d) }$,
replace $\Lambda (d)$ by (unknown yet) $d_{pk}$ as
\begin{equation}
\frac {\Lambda_0} {d_{pk} }=
\frac {1} {4 + ( l_0 / d_{pk} )^2 } +
\frac 3 4 \left( \frac {d_{pk}} {l_0} \right)
ln \left( 1 + \frac {l_0} {d_{pk}} \right)
\tag{6.6}
\end{equation}
and define a new fundamental spatial scale, $\lambda_N$,
directly related to $d_{pk}$,
since $d_{pk} = O ( \lambda_N )$:
\begin{equation}
\lambda_N = \sqrt {\Lambda_0 l_0} =
\frac { ( 3 / 8 \pi  )^{1/6}}   {\sqrt { \alpha }} N_e^{-1/3} 
\approx 8.2 \times N_e^{-1/3}
\tag{6.7}
\end{equation}
As opposed to $\Lambda_0$,
it does not depend on 
free electron path $l_0$ and 
therefore on temperature,
same as the Fermi wave vector, $k_F$.
(While $d_{pk}$ still depends on $l_0$,
this dependence is logarithmically weak at
$l_0 / \lambda_N \gg 1$.)
Using this scale,
we introduce now dimensionless variables
\begin{equation}
\eta = \frac {l_0} {\lambda_N }=
\sqrt { \frac {l_0} {\Lambda_0 }} =
0.122 \times N_e^{1/3} l_0 \ \ and \ \
\zeta = \frac {l_0} { d_{pk}}
\tag{6.8}
\end{equation}
where $\eta$ is a "free electron figure of merit"
($\eta \gg 1$ for either good metals, as e. g. for $Ag$,
or long mean free path of electrons, as e. g. for $Bi$),
and $\zeta$ is an inverse position
of peak absorption weighted by $l_0$.
Eq. (6.6) is rewritten then in the form
\begin{equation}
\eta^2 = 4 \zeta^2   /
[  3    ln ( 1 + \zeta ) + \zeta  / ( 1 + \zeta^2 / 4 ) ]
\tag{6.9}
\end{equation}
It makes it convenient to plot and analyze
figure of merit, $\eta$, $vs$
dimensionless peak position,
$s_{pk} \equiv d_{pk} / \lambda_N = \eta / \zeta$.
For poor conductors, we have
$\eta , \zeta \ll 1$ (i. e. $l_0 / \lambda_N \ll 1$),
the solution of (6.9) is $\zeta = \eta^2$, or
$l_0^2 / \lambda_N^2 =
l_0 / d_{pk}$,
and since $\lambda_N^2 = \Lambda_0 l_0$,
we have $d_{pk} = \Lambda_0$,
as expected when $\sigma = \sigma_0$,
i. e. the absorption peak position coincides
with that predicted by a simple theory
within which $\sigma = const = \sigma_0$,
In general, to find $d_{pk} / \lambda_N$,
for a given $\mu$, we need to
inversely solve (6.9) for $\zeta ( \eta )$.
This is greatly simplified for
good metals, whereby $\eta , \zeta \gg 1$ and 
$s_{pk} = O (1)$.
Eq. (6.9) then is reduced to 
$s_{pk} \approx 2 / \sqrt { 3 ln ( \eta / s_{pk} )}$,
and a good estimate can be obtained
via fast converging iterations, whereby $s_0 = 1$,
and $ s_n =  2 / \sqrt { 3 ln ( \eta /  s_{n-1} ) }$,
by e. g. using $n=2$ or even  $n=1$.

Eq. (6.5) allows to generate plots of reflectivity $R$,
transmittivity $P$, and energy losses $Q = 1 - R - P$,
$vs$ the thickness $d$ using (3.7)
[having in mind now that
for any given $d$ the parameter $\Lambda = \Lambda (d)$
depends now on $d$, via $\Lambda_0 / \Lambda (d)$  in (6.5)].
Fig. 3a, shows those plots
for the example of silver film [4].
One can see that the absorption, $Q$,
has a peak, $Q_{pk} = 0.5$ (and $R = P = 0.25$),
at the thickness $d_{pk} \approx 12.6$  {\AA},
as predicted by (6.9) (or simplified calculations
for $s_{pk}$, see the preceding paragraph),
based on the known data (see Table 1)
that the characteristic length 
$\Lambda_0 = c / ( 2 \pi \sigma_0 ) $ 
for silver is $\Lambda_0 \approx 0.84$ {\AA},
and the mean free path of electrons is $l_0 = 533$ {\AA}.
Based on those two numbers, 
we also estimate the new spatial scale as 
$\lambda_N = \sqrt { \Lambda_0 l_0 } 
\approx 21.2$  {\AA},
and the silver figure of merit as 
$\eta = \sqrt {l_0 / \Lambda_0} \approx 25.2$.
\begin{table} [ht]
\includegraphics[width=0.47\textwidth, height=0.3\textheight, angle=0]{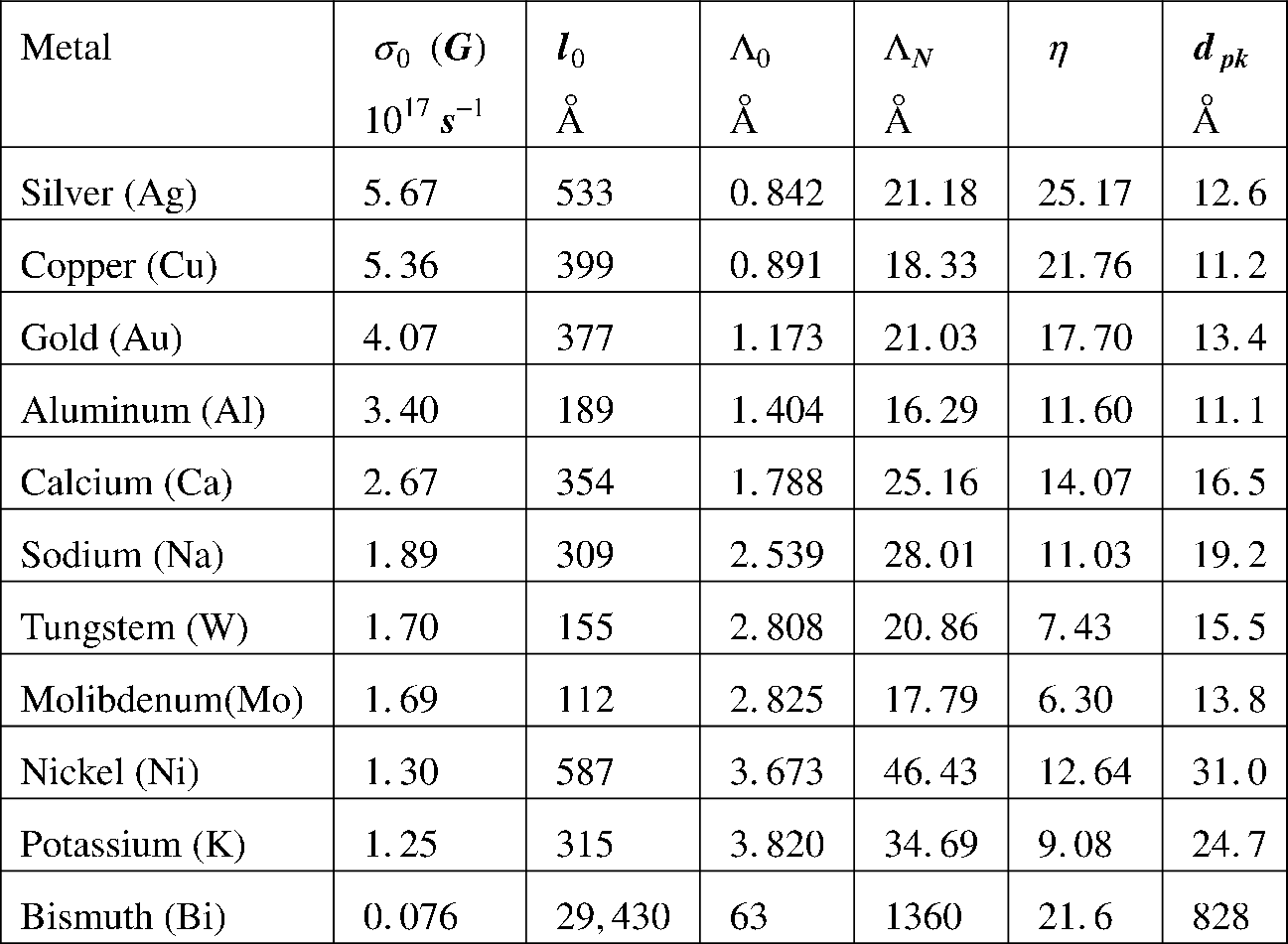}
\caption{Bulk conductivity $\sigma_0$, mean free path of
electrons, $l_0$, characteristic scales $\Lambda_0$
and $\Lambda_N$, 
free electrons figure of merit $\eta$, and 
peak thickness $d_{pk}$ for various metals.
The data for $l_0$ for the first 
10 metals are due to [50], for $Bi$ - to [52].}
\label{Table 1}
\end{table}
As one can see from Table 1,
good metals ($Ag$, $Cu$, $Au$, and $Al$)
have their conductivity $\sigma_0$ 
of the same order, which is also true
for their mean free path of electrons,
$l_0  \sim  200 - 500$ {\AA},
and the scales $\Lambda_0  \sim  1$ {\AA},
and $\Lambda_N  \sim  20$ {\AA}, 
resulting in $d_{pk}  \sim  12$ {\AA}.
The theoretically calculated
reflectivity $R$, transmittivity $P$,
and absorption $Q$
$vs$ the thickness $d$ for a silver layer
are shown in Fig. 1.

In view of the results of Section 5,
it is important to evaluate how
the major characteristics
of thin films changed for the
input/output environment different from free propagation,
e. g. when $n_1 + n_2 \not= 2$,
where $n_1$ and $n_2$ are
the refraction coefficients of
input and out media respectively.
Having in mind that due to (5.13),
$d_{pk}= \Lambda ( d_{pk}) ( n_1 + n_2 ) / 2$,
where $d_{pk}$ is the thickness
that corresponds to peak absorption, 
$Q_{pk}$ (5.13),
the ratio $\zeta = l_0 / d_{pk}$
is determined now by eq. (6.9) 
modified as
\begin{equation}
\zeta  = \eta 
\sqrt {  3    ln ( 1 + \zeta ) + 
\zeta  / ( 1 + \zeta^2 / 4 ) } / 
\sqrt{2 ( n_1 + n_2 )}
\tag{6.10}
\end{equation}
%
for good metals, $\zeta \gg 1$,
it can be further simplified as
$\zeta  \approx \eta 
\sqrt {  3    ln (\zeta ) / 2 ( n_1 + n_2 )}$.
Notice that $\zeta$
depends only on the sum $n_1 + n_2$,
and not on individual indices $n_i$'s,
while the optical characteristics
$Q_{pk}$, (5.13), $R$, and $P$ (5.14),
depend on $n_i$'s separately.
At that, if $ n_1 = n_2 \not= 1$,
we still have $Q_{pk} = 1/2$,
and  $R = P = 1/4$, as in free propagation.
Similarly to (6.9),
eq. (6.10) is readily solved numerically
by fast converging iterations,
starting with $\zeta_0 = \eta$.
For example for Aluminum
at $n_1 = 1$, $n_2 = 1.5$ 
(air+glass) we have $d_{pk} = 12.6 \AA$,
while at $n_1 = 1$, $n_2 = 4$
(air+silicon), $d_{pk} = 18.7 {\AA}$,
with $Q_{pk} = 0.8$.
\vspace{-.2in}
\section*{7. Experimental observation}
\vspace{-.1in}
\noindent
One of recent publications on experimental
measurements of the effect and observation of the peak
absorption in $nm$ films was the work by Andreev
and co-workers [11], who studied the optical
properties ($R$, $P$, and $Q$) of thin aluminum film
using radiation with $\lambda = 8 \ mm$.
Their results are depicted in Fig. 3
[11],  for a $Al$ film deposited on a glass substrate
with refractive index $n = 1.5$;
Fig. 3a is for the configuration whereby the wave is incident 
upon $Al$ film from air, and Fig. 3b - from the substrate.
(In those plots, the reflectivity is denoted by $R$, i. e. 
the same as in this paper, whereas $T$ denotes transmission,
i. e. corresponds to $P$ in this paper,
and $A$ -- absorption corresponding to $Q$ here.)
Theoretical plots were calculated
for the film environment consisting
of two different materials (air and glass),
using formulas similar to (5.10)-(5.12),
and the size-dependent conductivity
-- using equations similar to (6.5)
in the limit $d \ll l_0$.
As one can see, the experiment shows
a great qualitative agreement with the theory,
which is also true for quantitative agreement at
the thicknesses $d$ greater
than $20$ {\AA} (2 \ nm ).
However the position $d_{pk}$ of the maximum absorption
is almost double of that predicted
by the theory; the authors' explanation of that 
is that at that thickness ($1 - 2 \ nm$),
a thin films undergoes a structural transformation,
whereby it gets granulated
(which may depend very much
on the way the film was prepared [37])
or even breaks up into islands,
which results in much faster reduction
of averaged conductivity;
hence the shift of peak of absorption
to a greater thickness. 
We also note that in strongly 
granulated films, the "absorption"
calculated as $Q = 1 - P - R$,
could be very much due to strong scattering
[37] and not due to real losses in the film.
\begin{figure} [ht]
\begin{center}
\includegraphics[width=7cm, height=10cm, angle=0]{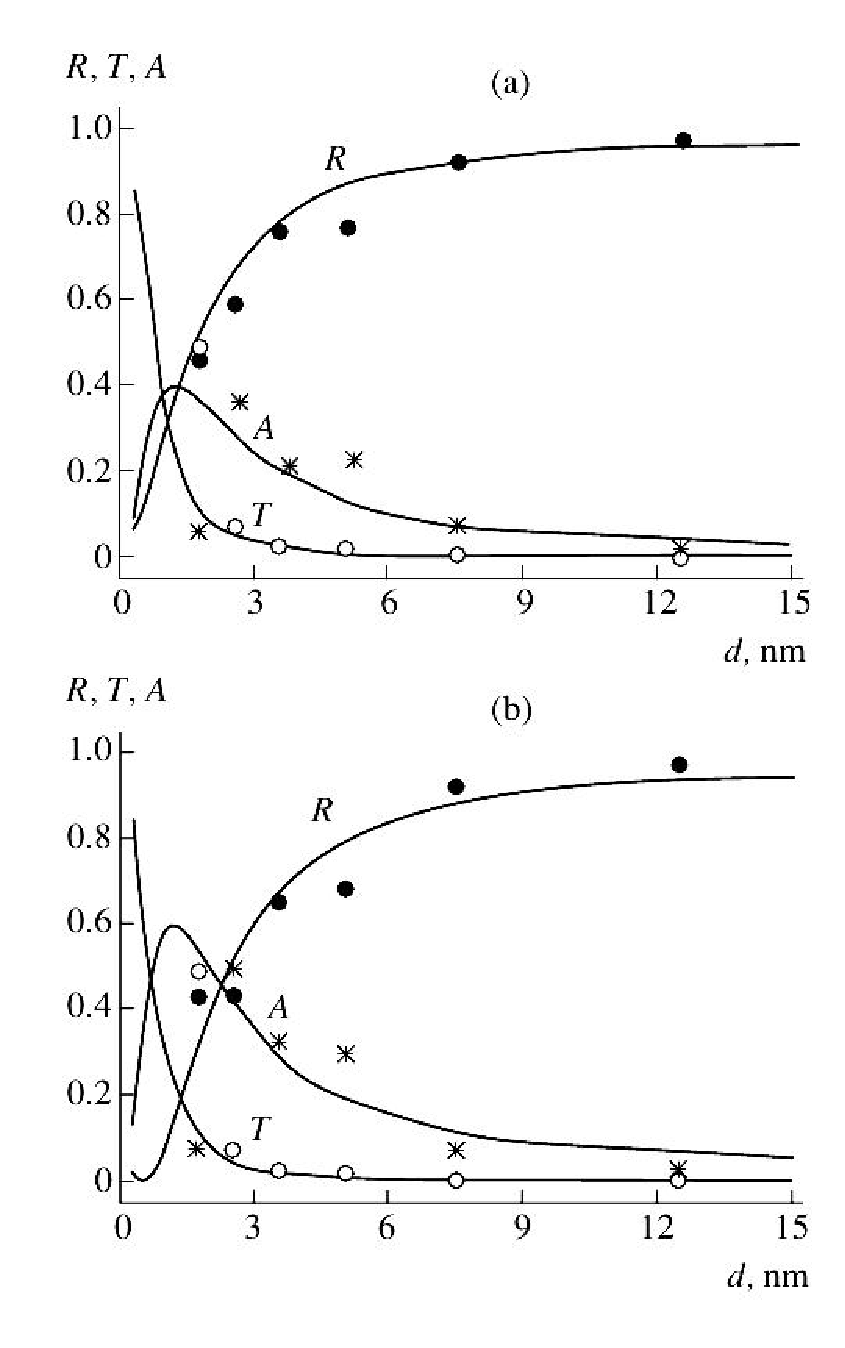}
\vspace{-0.2in}
\caption{Experimental and theoretical data [11]
for reflectivity, transmittivity,
and absorption of aluminum film $vs$ its thickness (in $nm$).
}
\label{fig3}
\end{center}
\vspace{-0.4in}
\end{figure}
Having in mind possible future experiments
to find out whether the theory based
on the assumption of a homogeneous layer
is still good around the peak of absorption,
one might be interested to use a metal with
{\it lower} intrinsic bulk conductivity
and thus - greater $d_{pk}$,
hence still relatively unperturbed structure
with greater number of atomic layers,
One can see from the Table 1,
for the metals with lower bulk conductivity 
($Ca$, $Na$, $W$, $Mo$, $Ni$ and $K$)
but roughly the same $l_0$,
the scale $\Lambda_N$ and peak position $d_{pk}$
predictably increase, up to $\sim  45$ {\AA}
and $\sim  31$ {\AA} (for $Ni$) respectively;
this should make it easier to measure all
the related effects in more homogenius structure.

Semi-metals such as e. g. tin, graphite,
bismuth, telluride, and their chemical compounds
(including most recently developed semi-metal polymers [53])
may be even more promising potential
candidates for further explorations of the
phenomenon considered here,
for they could have much longer mean free
path of electrons, and may provide
an arena of almost ideal homogeneous
(i. e. granulation-free) layers
that can be much easier to use for
more cleaner experiments.
A good example  is Bismuth ($Bi$) that has 
$\sigma_{Bi}^G \approx 0.76 \times 10^{16} s^{-1}$
[52] (see Table 1).
Having a mean free path
of electrons about $3 \mu m$ at $T \sim 300^o K$
it would exhibit impedance-math
absorption close to 50\% at the thickness near $80 \ nm$,
which allows to have it as a free-standing films.
At thicknesses below $20 - 30 \ nm$ and low temperatures,
$Bi$ becomes a semiconductor [54],
which would make it a different 
and even more interesting game.
%
%
\vspace{-.3in}
\section*{8. Free-space terminator and
coherent broadband interferometry}
\vspace{-.1in}
\noindent
In waveguides or transmission lines, 
the full absorption of incident wave 
is attained by a terminator whose
impedance matches that of the waveguide - but not in free space.
However, it was demonstrated in [5],
that such "black-body" (BB) can be realized
by using a thin metal layer of exactly
the thickness $d_{pk}$  
(which has only half of impedance of free space)
in a Sagnac interferometer.
It would then provide 100\% absorption (hence zero reflection) 
for the entire spectrum of incident radiation in one position 
and almost full transparency in another;
such a device might be of great interest to many applications.
The effect is due to ideal coherence between
incident and transmitted radiation 
for all the frequencies involved:
because of tremendously low distance of propagation the 
phase of transmitted wave is exactly the same as that of incident wave,
while the reflected wave has an opposite phase,
which is true for the entire spectrum.

To realize this effect one needs 
$counter-propagating$ waves of the same amplitude,
running normally to the layer (without the layer they 
would form a standing wave).
Since the amplitude reflection coefficient, 
the reflection of a straight-propagating (``+'') incident wave
of the unity amplitude at $d= d_{pk}$ 
will form a back-propagating wave, 
$E_{refl}^{(+)}$ with the amplitude $-0.5$.
At the same time, if a back-propagating 
(``-'') incident wave
have exactly the same phase at 
the film as the ``+'' incident wave,
its transmitted portion, 
$E_{trans}^{(-)}$ will have the same phase, 
and $p=0.5$, so that $E_{refl}^{(+)}=-E_{trans}^{(-)}$, 
and similarly, $E_{refl}^{(-)}=-E_{trans}^{(+)}$.
Thus, there will be $no$ waves escaping
from the film into any direction, and {\it the energy of
both waves will be fully absorbed}! 

In such a case, the layer is to be located in
the anti-node (i. e. maximum of electric field) 
of the original standing wave,
so one should expect the largest 
absorption due to largest
generated electrical current. 
It is clear then that when 
the film is located at the node 
of the original standing wave, where the electrical
field vanishes, 
{\it the absorption vanishes too}, 
as if there is no absorbing layer at all. 
\begin{figure} [htb]
\vspace{-.1in}
\begin{center}
\includegraphics[width=5.6cm, height=6cm, trim=0 4cm 0 0, clip, angle=0]{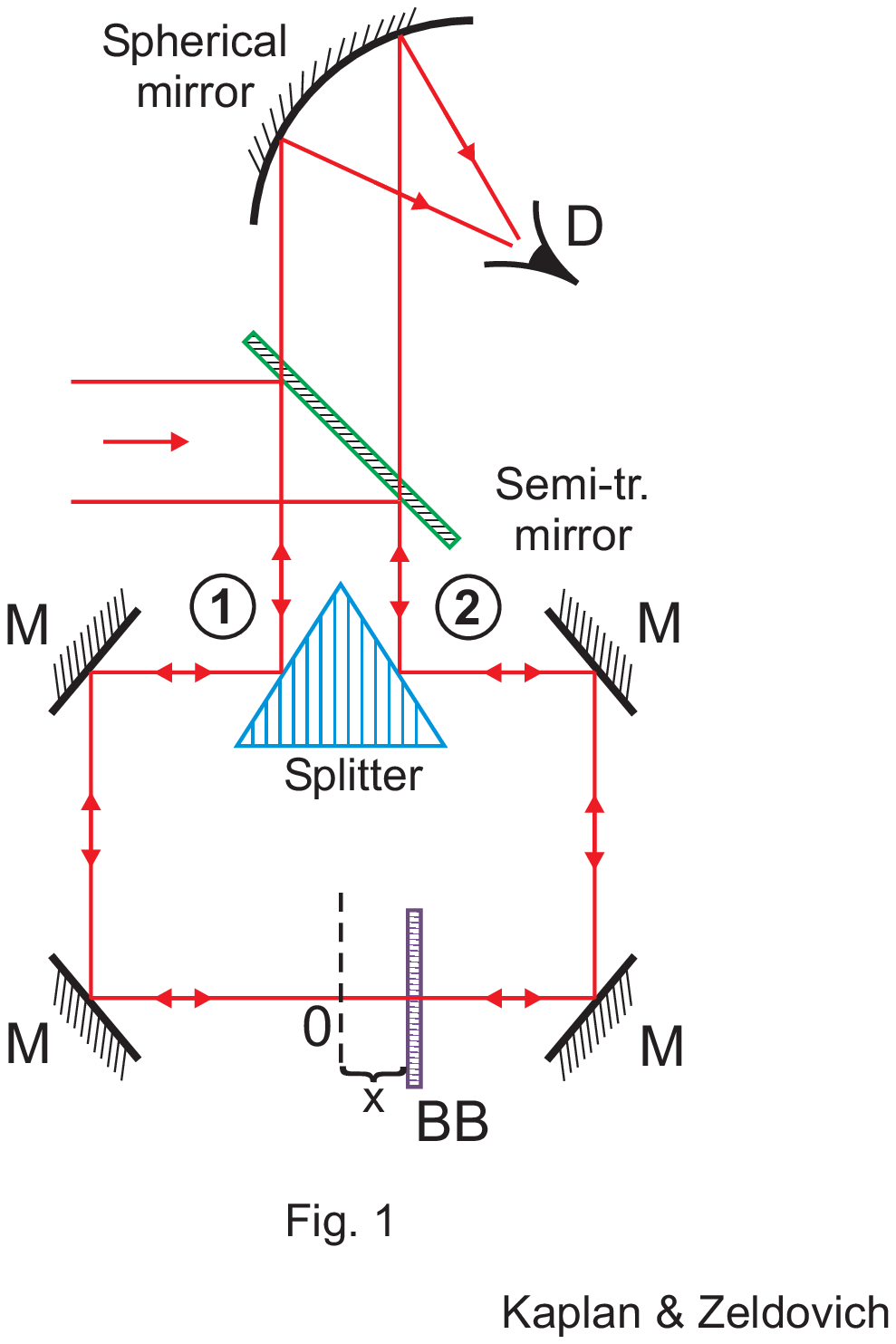}
\caption{Ring (Sagnac) interferometer using a black-body element (thin metallic film) BB.
Notations: M - metallic mirrors, D - intensity detector.}
\label{fig4}
\end{center}
\vspace{-.1in}
\end{figure}
Such a system can be realized as a ring 
(Sagnac) interferometer, Fig. 4, 
in which an incident wave
is split into two waves of equal intensity.
Those are then made to propagate
against each other on the path
with a thin metallic film in the middle.
If the film is positioned at 
the anti-node of the standing wave, 
the light energy will be fully absorbed, 
whereas if it coincides with a node,
there will be a full reflection. 
For an $\omega$-monochromatic wave 
the BB-reflection, $R^{(BB)}$ normalized 
to that of the system
{\it without the BB layer}, $vs$ offset $x$ of the layer
from the ant-node is 
$R^{(BB)}=\sin^2 (xk)= \sin^2 ( \omega \tau / 2 )$
where $\tau=2 x / c$ and $k=\omega /c$.
For an incident radiation 
with arbitrary temporal profile, $E ( t )$
with $\int_{-\infty}^{\infty} E ( t ) d t=0$,  
and normalized autocorrelation function,
$F ( \tau )= < E ( t ) E ( t-\tau ) > / < E^2 ( t ) >$,
where brackets $<\ >$ stand for time averaging,
$< \zeta >= \int_{-t_{av}}^{ t_{av}} \zeta dt/2 t_{av}$ as $t_{av} \rightarrow \infty $ 
($F ( \tau ) = F ( -\tau )$), the respective  spectrum is:
\begin{equation}
S ( \omega )= 
\frac{1}{ \pi } \int_{0}^{\infty} F ( \tau ) \cos ( \omega \tau ) d \tau.
\tag{8.1}
\end{equation}
If one of the output channels, 
e. g. channel 2, behind the 
film, Fig. 4, is blocked,
the output signal in channel 1, 
$E_{1out}$, is formed then by the input 2, 
$E_{2in}$, that gets through the entire 
loop without change of sign and attenuated by
the factor of $2$, and by the input 1, 
$E_{1in}=E_{2in}$, that gets reflected 
by thin film with the same attenuation,
but with the change of sign, so that
\begin{equation}
E_{1out} ( \tau ) \propto E_{in} ( \tilde{t} )- E_{in} ( \tilde{t} - \tau )
\tag{8.2}
\end{equation}
where $\tilde{t}= t- t_0$ is a retarded time, with $t_0 = L / c$ is a full time delay of
the light to go around the full ring of the length $L$.
The normalized BB-reflection at the fixed delay $\tau$ is then:
$R_1^{(BB)} ( \tau )= {< E_{1out}^2  ( \tau ) >} / {4 < E_{in}^2  ( 0 ) >}$.
Using (8.2), and having in mind that due to (8.1), 
$F ( \tau )= \int_{-\infty}^{\infty} S ( \omega ) 
e^{i \omega \tau } d \omega=$ $2 \int_0^{\infty} S ( \omega )  \cos ( \omega \tau ) d \omega$, and
$F ( 0 )=1=$ $2 \int_0^{\infty} S ( \omega ) d \omega$,
we have 
%
\begin{equation}
R_1^{(BB)} ( \tau )= \frac{ 1 - F ( \tau ) }{2}=2\int_0^{\infty} S ( \omega ) \sin^2
 \left( \frac{\omega \tau}{2} \right) d \omega
\tag{8.3}
\end{equation}
Note that $R_1^{(BB)} ( 0 )=0$, $R_1^{(BB)} ( \tau )=O ( \tau^2 )$
as $\tau \rightarrow 0$, and, for $F ( \infty ) = 0$, we have $R_1^{(BB)} ( \infty )=1/2$.
A typical example is a Gaussian spectrum, $S ( \omega )$, with an arbitrary
bandwidth $\Delta \omega$ centered around some frequency $\omega_0$:
\begin{equation}
S ( \omega )= [ exp ( - s_+^2 ) + exp ( - s_-^2 ) ] 
( 2 \Delta \omega \sqrt \pi )^{-1}
\tag{8.4}
\end{equation}
where $s_{\pm} = ( \omega \pm \omega_0 ) / \Delta \omega$.
The total reflectivity is then
\begin{equation}
R_1^{(BB)} ( \tau )= \left[ 1 - \cos ( \omega_0 \tau ) X \right] / 2
\tag{8.5}
\end{equation}
with $X = \exp \left[  - ( \tau  \Delta \omega / 2 )^2 \right] $.
Fig. 5 depicts $R_1$ $vs$ $x$ for various ratios $\Delta \omega / \omega_0$.
\begin{figure} [htb]
\includegraphics[width=4cm, height=6.5cm, angle=270]{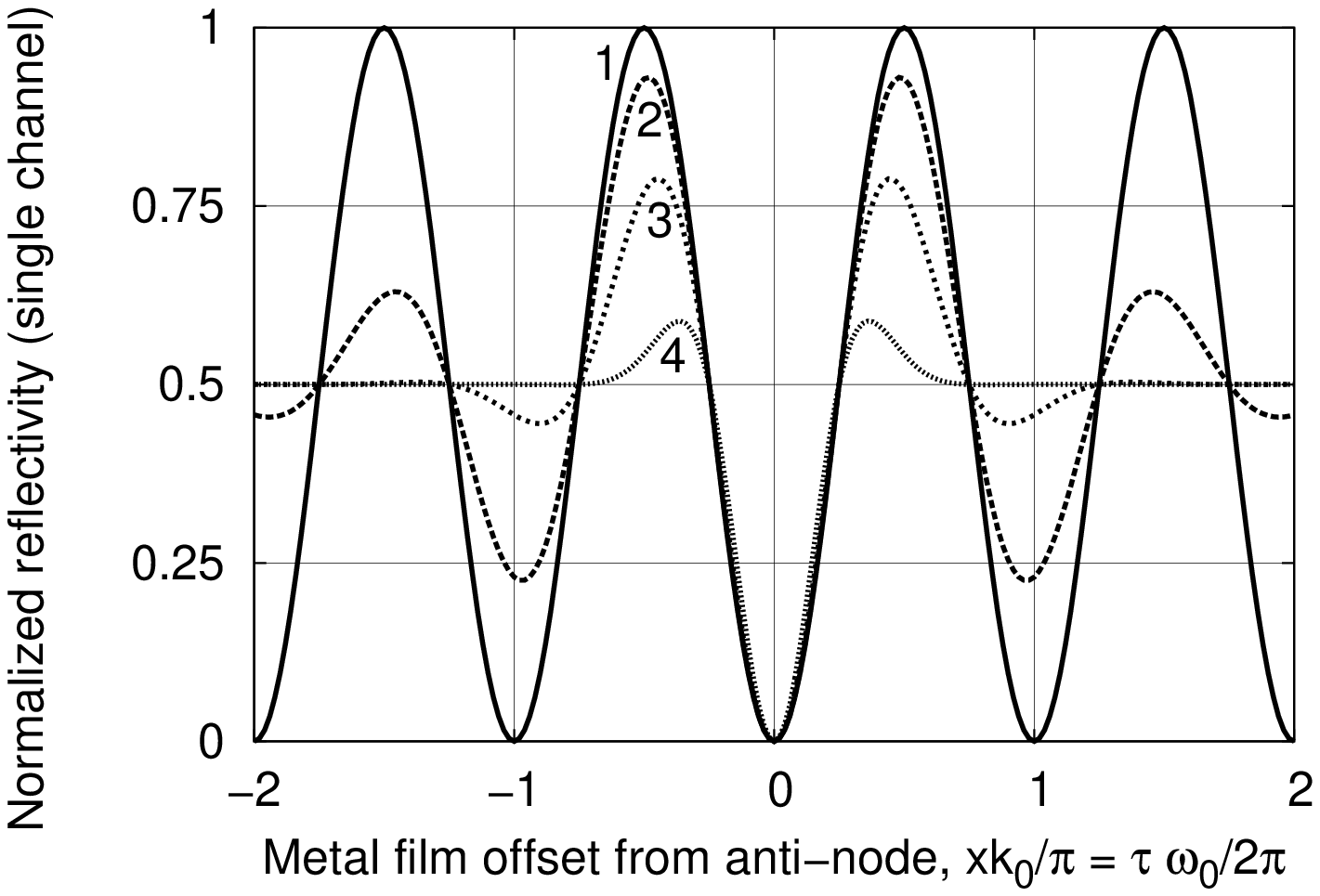}
\begin{center}
\vspace{-.2in}
\caption{Normalized reflectivity $R_1^{BB}$ of 
the black-body interferometer $vs$ normalized offset
of metal film, $x k_0 / \pi=\tau \omega_0 / 2 \pi$,
from $x= 0$,
for the broad-spectrum signal Eq. (8.5), for a single channel.
Curves: 1 - $\Delta \omega = 0$ (monochromatic wave);
2 - $\Delta \omega = \omega_0 / 4$,
3 - $\Delta \omega = \omega_0 / 2$,
4 - $\Delta \omega=\omega_0 $. }
\label{fig.5}
\end{center}
\vspace{-0.2in}
\end{figure}
With both of the output 
channels opened, the full output signal is
$E_{out} ( \tau ) \propto E_{in} ( \tilde{t} )$ $ -  ( 1 / 2 )   [ E_{in} 
( \tilde{t} - \tau ) + E_{in} ( \tilde{t} + \tau ) ]$
similarly to (8.3), the full BB-reflection is
$R_\Sigma^{(BB)} ( \tau )= [ 3 - 4 F ( \tau ) + F ( 2 \tau ) ] / 8$
or
\begin{equation}
R_\Sigma^{(BB)} ( \tau )= 
2 \int_0^{\infty} S ( \omega ) 
\sin^4 \left( \frac{\omega \tau}{2} \right) d \omega
\tag{8.6}
\end{equation}
In the case of the Gaussian spectrum (6) we have:
\begin{equation}
R_\Sigma^{(BB)} ( \tau )= [  3 - 4 \cos ( \omega_0 \tau ) X +
\cos ( 2 \omega_0  \tau )  X^2 ] / 8
\tag{8.7}
\end{equation}
For monochromatic input, $\Delta \omega =0$, we have
$R_\Sigma^{(BB)} =\sin^4 ( \omega_0 \tau / 2)$.
In the case of white-like noise. i. e. 
there is no distinct central frequency of the signal,
$\Delta \omega \gg \omega_0$, or simply $\omega_0=0$,
equations (8.5) and (8.7) are reduced to
\begin{equation}
R_1 = ( 1-X ) / 2 ; \ \ \ R_\Sigma = {R_1^2(3+2X+X^2)}/{2},
\tag{8.8}
\end{equation}
At $\tau \Delta \omega \equiv q \ll 1$,
we have $R_1^{(BB)} ( \tau ) \approx q^2 /8$ 
while $R_\Sigma^{(BB)} ( \tau ) \approx 3 q^4 / 64$,
i. e. double channel detection is much more sensitive 
to the high-frequency details of the spectrum.
In general, using both reflectivities, (8.5) and (8.7), 
one may substantially enhance the temporal \& spectral resolution,
because of simultaneous auto-correlation at 
two different delay times, $\tau$ and $2 \tau$.
Fig. 5 depicts $R_1$ $vs$ $\omega_0 \tau$ for 
various ratios $ \Delta \omega / \omega_0$.
\vspace{-.1in}
\section*{9. Applications and Outlook}
\noindent
Having a nanometer-thick film
absorbing 50\% or even 100\%
of incident power of extremely broad
spectrum may have quite 
a few promising applications.
We will discuss a few of them,
yet there is no doubt that there could be others.
Besides, one can expect
some interesting directions of research
related to such films.

So far we discussed a coherent spectroscopy of
signals with super-broad,
almost white-noise spectrum.
To measure such signals most 
of the elements of the Sagnac interferometer
must be metallic, including all the mirrors, 
and the semi-transparent mirror should be made also
the same way as the black-body element, 
i. e. by using again a very thin metallic layer.
This is necessary to extinguish 
any possible resonant or frequency-sensitive
effects if the mirrors are made of dielectric layers.
This kind of spectroscopy would be appropriate
for  sub-visible domain down to mid-infrared.
It is well suited for Terahertz technology; 
other applications may include the detection
of high-frequency coherent features that may 
allow for detecting an information transmission
in "pseudo-white-noise" signal, and potentially, 
in $mw$ radiation from the space that may 
be helpful in the detecting extraterrestrial signals, 
as well as in low-level signal such as primordial thermal radiation.
In all these potential applications the important factor is that in contrast
to regular auto-correlation techniques, 
whereby the auto-correlation signal at small
delay times $\tau$ is finite, the BB-interferometry 
produces a zero output at $\tau=0$, 
which may greatly increase its sensitivity 
compared to that of a regular auto-correlation.
For certain applications, e. g. for primordial radiation, 
special care should be taken of black-body 
radiation of the BB-element (same as the other mirrors) by
cooling it down with e. g. liquid helium. 
%

Further modification and enhancement of the 
BB-interferometry, especially for narrow-band
signal, may be attained by employing more than 
one metallic layer and using ensuing resonances.
For example, for the monochromatic radiation 
with wavelength $\lambda$, if the
spacing between layers with $Q=1/2$ is 
$\lambda / 2$, the system is fully transparent,
if irradiated from both directions.
Inversely, a reflection resonance would  
exist if the spacing is $\lambda / 4$.
In this case, the amplitude of reflection 
of each of the counter-propagating waves
is $r=- 2 / 5$, and transmission, $t=1/5$.
If the couple is positioned strictly at the center of the ring 
interferometer, the 
amplitude reflection for each of the waves is $-1/5$,
and the intensity reflectivity is thus $4$\%.
This is not far from total zero as with a single layer, 
but the system has substantial selectivity to the frequency.
In thin-metal multi-layered structure [20,21]  
the resonant effect will be enhanced.

Another feasible application might be related
to the use of $Q = 0.5$ layers
for detecting and imaging/visualization 
of $IR$ on $mw$ radiation
by covering a metallic film with thermoluminescent layer 
(i. e. whose luminescence strongly depends on temperature)
continuing along the line of the original 
research [14-16] but using more advanced
materials (see e. g. [55]).
If such a material is preliminary 
irradiated by e. g. UV-radiation
and then --  by infrared,
the spots where infrared is stronger, will be 
heated up enough to trigger visible thermoluminescence
from such spots.  The visible optical image 
is expected to have then very high
spatial resolution, since the heat transfer
{\it along the layer} would be negligible
due to its extremely small thickness.

Another expected effect is
related to nonlinearity of the layers 
slightly thinner than the 
peak absorption, e. g. less than $1 - 2$ $nm$.
At the thicknesses corresponding to the formation
of isolated "islands" of metal,
which are still close to each other,
the local field due to formation of 
plasmons can get enhanced by orders
of magnitude and due to closely packed
islands induce tunnelling transitions and discharges,
i. e. strongly nonlinear effects that may
result in high harmonics generation.

It would be of great application interest
to develop a tool of fast and efficient
modulation of optical properties
of ultra-thin films, especially in
the vicinity of maximal absorption,
by using an electro-optical
effect to control behavior of free electrons
in a film, as e. g. in [56].

Application-wise it would be interesting to
use inexpensive "artificial" 
metal-like polymers, i. e. highly conductive
doped polyacetone whose electrical conductivity
can be varied over the range of eleven orders
of magnitude [57], polyanilin [58],
and others (see review [59].
Related to that would be 
development of controllably produced
2D spatial modulation of the 
conductivity of the thin film,
allowing thus opportunity
to design 2D photonic nano-crystals [60,61]
with easily designable patterns.

As we have already discussed earlier,
semi-metals present an interesting opportunity
to study impedance-matching
films, as most of them have a 
a very long mean free path of electrons;
$Bi$ would be probably the most promising.
Actually, it was the element which
was first predicted [62] and experimentally observed [63]
to show quantum-size effect at low temperature. 
In that effect, when the 
film thickness becomes comparable 
with the effective de Broglie wavelength
of electron, it would exhibit
oscillations of its properties $vs$ e. g. its thickness.
It would be of fundamental interest
to explore possibility of time-dependent analogy
of this effect in phase-matching $Bi$
films under modulation of the incident radiation.
\vspace{-.2in}
\section*{Conclusion}
\vspace{-.1in}
\noindent
In conclusion, we reviewed major features 
of the frequency independent reflection of the radiation 
from ultra-thin metallic layers
within large, so called sub-visible
domain from the $rf$ to $mw$ to $mm$ to $mid-infrared$,
or even $infrared$.
We demonstrated a very universal
optical properties of such layers:
they remain almost ideally reflectant 
(and almost non-absorbing)
at the thicknesses orders of magnitude
shorter that skin-layer at any frequency,
down to a certain depth scale, typically a few $nm$,
which depends only on the
number density of free electrons.
Near that scale the optical 
parameters undergo dramatic change, 
whereby the reflectivity becomes equal 
to the transmittivity (25\% ),
while 50\% of the incident energy is absorbed
(under certain arrangement the absorption can go up to 100\%).
From the general EM point of view, this situation 
corresponds to a layer's wave impedance matching exactly 
half of the impedance of free space.
A major role in this scale formation
is played by the size-affected conductivity
directly related to the mean free electron path
being "clipped" by the walls of the film.
We also considered arbitrary
environment (metal film sendwiched
between dielectrics with different
refractive indeces).
We pointed out quite a few feasible
applications of the phenomenon 
and related research direction.
\vspace{-.2in}
\section*{Acknowledgments}
\vspace{-.1in}
\noindent
The need to have this review 
became obvious during phone conversation
with Prof. Eli Yablonovitch 
a while ago regarding the results of Ref. [4],
and the author gratefully acknowledges
that discussion.
Part of the work was done 
at Weizmann Inst. of Science, Israel,
and the author is grateful to G. Kurizki, I. Averbukh, Y.
Prior, and E. Pollak of Weizmann Inst.
for their kind hospitality during his
stay there as a visiting professor.
\vspace{-.1in}
\section*{Appendix A:  Maxwell Equations in Gaussian Units}
\vspace{-.1in}
\noindent
We used here
the Maxwell equations in Gaussian units,
and Drude model for metal as
a gas of quasi-free electrons :
\begin{equation}
\nabla  \times  \tilde{\vec{E}}   = 
-  \frac 1 c  \frac {\partial \tilde{\vec{H}}}  {\partial t} ;
\ \ \ \
\nabla  \times  \tilde{\vec{H}}   =   \frac 1 c  
\frac {\partial \tilde{\vec{E}}}  {\partial t}  +  
\frac {4 \pi \tilde{\vec{j}}}  c ;
\tag{A.1}
\end{equation}
where $\tilde{\vec{E}}$ and $\tilde{\vec{H}}$ are
respectively electrical and magnetic fields,
$\tilde{\vec{j}} = \sigma  \tilde{\vec{E}}$
is current density,
and $\sigma$ is a $dc$ conductivity
of a metallic layer; 
$\sigma = 0$ outside the layer.
(The Drude model is quite adequate
model of conductivity
for optical properties of metals in sub-visible
domain, while their thermal properties
are not considered here.)
In the Gaussian units
$\sigma$ is measured by the same unit as frequency,
i. e. $[ \sigma ] = s^{-1}$.
The SI units for the conductivity $\sigma$,
[ $( \Omega \cdot m )^{-1}$ ], 
(or resistivity $\rho = 1 / \sigma$)
used often in the literature,
can be conversed to the Gaussian units as
\begin{equation}
{\sigma^G} / {\sigma^{SI}} =
{\rho^{SI}} / {\rho^G }
\approx 9 \times 10^9 \ \Omega  
\ {\cdot m} /   s
\tag{A.2}
\end{equation}
For $\omega$-monochromatic radiation,
we represent, as usual, any field, $\tilde{\vec{F}} ( \vec{r} , t )$,
as a product of time-independent amplitude, 
$\vec{F} (  \vec{r} )$,  and time-dependant exponents
$\tilde{\vec{F}} ( \vec{r} , t ) = $
$(1/2)  \vec{F} ( \vec{r} ) e^{{-} i \omega t } + c. c.$,
and rewrite (A.1) as
\begin{equation}
\nabla  \times  \vec{E}  = i k  \vec{H} ;      
\ \ \ \
\nabla  \times  \vec{H}  = -  i \epsilon ( \omega ) k \vec{E} 
\tag{A.3}
\end{equation}
where $k = \omega / c = 2 \pi / \lambda$
and $\lambda = 2 \pi c / \omega$ are respectively
the wave-number and wavelength of the wave in a free space,
and $\epsilon$  is a dielectric constant; in 
free-space in Gaussian units we have  $\epsilon = 1$, and
inside the layer,
$ \epsilon = \epsilon_m = 1 + 4 i \pi \sigma / \omega $.
Under a "good metal" condition, $ | \epsilon_m  | \gg 1$,
or $\sigma \gg \omega$, $\epsilon_m$
can be well approximated by a purely imaginary quantity,
Eq. (2.2), where a scale $\Lambda$ [4],
was introduced in (1.6),
and skin depth $\delta$ is as defined in (1.2).
Dropping a "vacuum" term "1" in $\epsilon_m$
is equivalent to neglecting the term
$(1/c)  \partial \tilde{\vec{E}} / \partial t$ in 
the second equation in (A.1),
which, if we use current, $\vec{j}$, can be also
rewritten as a magneto-quasi-static equation:
\begin{equation}
\nabla  \times  \vec{H}   = 
Z_0 \vec{j} ; \ \ \ \ \ \     
Z_0 = 4 \pi / c
\tag{A.4}
\end{equation}
where $Z_0$ is the wave impedance of a free space 
in Gaussian units
(in $SI$ units $Z_0 = 120 \ \pi  \Omega \approx 377 \ \Omega$).
[It is worth noting that a plasma model of free electrons,
and related formula for dispersion
$\epsilon = 1 - \omega_{pl}^2 /
\omega ( \omega + i / \tau )$,
where $\omega_{pl} = \sqrt {4 \pi N_e e^2 / m}$ 
is plasma frequency,
while meaningful at higher frequencies, 
is of little use in the quasi-static case,
where the induced dynamics is much slower than
the relaxation time $\tau = l_0 / v_F$, 
where $l_0$ is a mean free path of electrons,
and $v_F =$ $\sqrt { 2 W_F / m}$ is a Fermi velocity,
see (6.1), $\omega \tau \ll 1$.
However, the above formula for $\epsilon_m$
is still consistent with plasma formula for
$\epsilon$ in the limit $\omega \tau \rightarrow 0$
having in mind relationship (6.1).]
\vspace{-.3in}
\section*{Appendix B. Side notes: 
metallic films, circa 1962-64}
\vspace{-.05in}
\noindent
In 1961, NASA launched first inflated balloon satellite, Echo 1,
to be used as a passive reflector of microwave radiation,
to detect the traces of atmosphere by monitoring
the de-acceleration of the balloon in time.
It was followed by much larger balloon satellite, Echo 2,
launched in 1964.
Both of them were made of thin mylar film
coated by a very thin aluminum foil
to facilitate a mirror-like reflection of the radiation.
In 1962, this author, fresh-graduated with his MS 
degree in general physics in 1961
with focus on "radiophysics",
worked at a government R\&D lab near Moscow,
that was developing inflated
balloons for meteorological and reconnaissance purposes
for Russian Air Force. 
He was asked to look into possible applications of
Echo-like satellites: the lab was looking
into the way to join rapidly growing space industry.

However, soon the idea of cat-copying Echo-satellite
was abandoned: the rocket-happy "big boys"
of Russian space industry apparently
were not much interested.
But he kept playing with the subject,
starting with the reflectivity of aluminum foil 
-- there were plenty of
aluminum-coated mylar films around, and he 
did some experimenting with them, trying to get voice-modulated
and electrostatic-controlled reflection
of large mirrors with a film stretched 
over a large rigid rim,
and a primitive yet 
efficient telescope: toys, basically.
His training called for the use of good theory;
as a warm-up exercise, he calculated
the reflection and transmission of thin metallic foils.
His seemingly straight expectation was that
in a $mw$ domain, a foil should start loosing its
reflectivity and increase its transparency,
when its thickness is just around the skin depth.
No such luck; to his great surprise, the reflectivity
kept staying close to 100\% even
when the foil thickness got 
orders of magnitude lower than that...
Greatly puzzled, he kept repeating 
his calcs -- with the same result...
All the sources available to him didn't
indicate anything like that either.
He finally showed his result to the lab bosses,
emphasizing that one can now reduce the weight 
of a potential satellite -- not a small feat those days.
He was met with derisive comments 
about his "elite-training".
\vspace{-.015in} 

He wrote his paper anyway, 
and it took more than a year to fight
reviewers off; it was accepted
then by a decision of a willful  editor in chief
(Prof. B. Z. Katsenelenbaum),
who checked out all the calcs by himself
(can you find an editor like that these days?...),
and published in 1964 [4].
The author even got a national award 
for "a best paper by a young scientist",
but it was meaningless for 
his further research career in Russia anyway,
especially considering his increasing involvement in
dissident human rights activity.
He never returned to the subject again
(till 2005, when Boris Zeldovich 
came up with a new twist about it,
and they published a paper [5] on the subject).
He's got his PhD on a completely 
different subject (high-order
subharmonics in nonlinear parametric oscillators)
on which he did his research for MS degree and
published it (as a sole author too),
even before the thin-film paper. 
Closer to the end of 60-ties he switched to 
lasers and nonlinear optics,
including predictions of self-bending effect,
and later on -- optical bistability 
and switching at nonlinear interfaces.
In 1979, carrying two suitcases and empty wallet,
he came as a refugee to the US,
where he immediately got back to his
research on nonlinear optics at MIT,
continued later on at Purdue and then Johns Hopkins,
in particular on nonlinear interfaces, hysteretic
relativistic resonances of a single cyclotron electron,
sub-femtosecond pulses, 
and shock waves in cluster explosions.
A whimsical but lucky part of all of that was 
that it was the Air Force (again) Office of 
Scientific Research, this time of the US, that
kept supporting him for 35 years;
his steadfastly encouraging and supportive 
program manager all that time was Dr. Howie Schlossberg,
while his diverse and forever shifting research interests
strayed far away from those early subjects.

\end{document}